\def\V{V_{606}}
\def\I{I_{814}}

\documentstyle[12pt,aaspp4,astrobib,psfig]{article}

\begin{document}

\title{Selection Effects and Robust Measures of Galaxy Evolution}

\author{Henry C. Ferguson}
\affil{Space Telescope Science Institute, Baltimore, MD 21218\\
ferguson@stsci.edu}


\begin{abstract}
A variety of subtle, and not-so-subtle selection effects influence the
interpretation of galaxy counts, sizes and redshift distributions in
the Hubble Deep Field. Comparison of the different HDF catalogs
available in the literature and on the world-wide-web reveals
generally good agreement, although the effects of different
isophotal thresholds and different splitting algorithms 
are readily apparent. As the basic source detection and photometry
algorithms are similar for the different catalogs, the selection
effects are likely to affect them all.

Through simulations, we explore the utility of image moments for
inferring the true sizes of galaxies. The truncation of galaxy profiles
at a fixed isophote has serious consequences, which limit constraints
on the size distribution to galaxies with isophotal magnitudes $\I < 27.5$.
Present-day $L^*$ spirals would be undetected in the HDF above
redshifts $z \approx 1.2$, and present-day ellipticals would disappear
at $z \approx 1.8$. 

The Lyman break provides a way to identify high-redshift
galaxies at very faint magnitudes. However, galaxies that are at
redshifts high enough to vanish from the HDF F300W or F450W filters
also suffer severely from photometric biases. For example, at fixed
total apparent magnitude and physical scale length, 
a galaxy at $z=4$ will have a mean 
surface brightness 1.2 mag fainter than a galaxy at $z=2.75$. 
This lower surface brightness will result in an apparent decrease in 
the number density of objects, and the inferred luminosity density, 
even for models where there is no intrinsic evolution.  
We illustrate the effects of these biases on the estimates of the 
number of Lyman ``dropouts'' in the HDF and on the luminosity density 
at $z > 2$.

\end{abstract}

\section*{Introduction}

In trying to decipher the origin of galaxies from a collection of fuzzy patches
on the sky, we must keep in mind that even the Hubble Deep Field, with
its exquisite depth and resolution, provides a distorted view of the
universe at large. This view is distorted by the fact that we are
looking only at optical wavelengths, which for the most part probe the
rest-frame ultraviolet portion of the spectra of the galaxies of
interest.  It is distorted by the fact that the background noise of the
detector limits detection of galaxies to those that exceed a certain
threshold over a certain number of pixels: faint extended objects are
extremely difficult to detect; galaxies with multiple peaks in their
light distribution may appear as separate objects.  It is distorted by the
fact that many of the galaxy images overlap, making it difficult to
separate one galaxy from the next. Finally, it is distorted by physical
effects such as obscuration by dust or gravitational lensing. 
These effects of dust and lensing are amply discussed in this conference by 
Madau, Meurer, Rowan-Robinson, Blandford, and others. I will focus on the
selection effects, seeking to explore and quantify some of the concerns that
have colored many of the discussions of the HDF and other deep galaxy surveys.

\section*{HDF Catalogs}
In making the HDF data set non-proprietary, one of the hopes was that
different groups would be stimulated to reduce and analyze the data
independently with different scientific objectives and different
algorithms. To some extent this has happened. The data have been largely
reprocessed by several different groups \cite{FGB96,Ratnatunga96}, and 
different techniques have been used for source detection and photometry 
\cite{WBDDFFGGHKLLMPPAH96,LYF96,ESG96,Couch96,MSCFG96}.
As much of the discussion at this meeting is focused on comparing
what we see in the HDF to what we expect from models of galaxy evolution,
it is worth examining some of these catalogs to see how well they
agree with each other, and how well they reproduce our own preconceptions
of how sources should be counted.

For this purpose, I have chosen four catalogs that have been used
in HDF publications. All the catalogs have in 
common the feature that they convolve the data to smooth the noise
over a scale relevant to the sources being sought, and then mark
as sources those objects that have a certain number of connected
pixels more than some multiple of the rms background fluctuations.
This procedure is equivalent to Wiener filtering with a Gaussian source
model, and
is optimal in the least-squares sense for detecting unresolved
objects against a random, uncorrelated background. It is not so clearly
optimal for finding galaxies in an image where galaxies have a wide
range of sizes and have isophotes that in many cases overlap. 
However, to my knowledge there have not been any HDF catalogs based,
for example, on wavelets or on median filtering over a variety of scales, which
might have a different set of selection effects.

\subsection*{Completeness}

The catalogs are briefly summarized in Table 1, which lists the 
software package used,  the number of objects found, and the rough
detection criteria. The total source
counts are influenced primarily by the isophotal threshold. To a certain
extent it is a matter of taste how deeply to push into the noise.
The \citeN{MSCFG96} catalog is the deepest (or least conservative)
while the \citeN{LYF96} catalog is the shallowest (or most
conservative). While the total number of sources varies by more than
a factor of two between the catalogs, down to $\V = 27.8$ the counts
in the different catalogs agree to within 20\% of the mean.

\subsection*{Object splitting and merging}

A variety of schemes have been developed for splitting up sources that
have multiple peaks above the detection isophote into separate
``parent'' and ``daughter'' objects. There is no mathematical rigor in
these techniques. Identification of daughter objects is typically done
by passing successively higher isophotal thresholds over the image and
making lists of sources within sources. Different algorithms are then
used to ``merge'' the various pieces back into objects that are likely
to be part of the same galaxy.  \citeN{WBDDFFGGHKLLMPPAH96} 
use color information to help with this merging; in other cases
the decisions were made using information from one band. 

The differences in how subcomponents are counted are illustrated in
the sections of the F606W image shown in Figs. 1-4, where we show
the positions of galaxies with total magnitudes brighter than 28.5
from the four different catalogs. Broadly speaking, the 
\citeN{LYF96} catalog tends to do the least splitting of objects
within common isophotes, while the \citeN{MSCFG96} catalog does the
most splitting. There are many cases, such as the peculiar galaxy
just to the right and above the center of the images, where it is
not at all obvious how one should count. Even more problematical
is the object that looks like a late-type spiral which is superimposed
on the outer isophotes of a bright elliptical galaxy near the bottom
of the frame. In the \citeN{LYF96} catalog, this object
is counted as part of the elliptical. In the \citeN{Couch96} catalog,
it is counted as four objects. It is counted as five by \citeN{WBDDFFGGHKLLMPPAH96}
and 4-6 by \citeN{MSCFG96}. 

From an analysis of the angular correlation function in their HDF 
catalog (constructed using DAOFIND), \citeN{CGOR97} 
suggest that such difficulties lead to an
overestimate in the number of faint galaxies by a factor of 2.5.
While this estimate strictly applies only for the subset of 196 color-selected
high-redshift candidate objects in their catalog, such overcounting
is likely to be true at some level for galaxies at all redshifts in 
all the catalogs. However, two important facts 
mitigate the effect of this problem on the interpretation 
of deep galaxy counts. The first is that the oversplitting of some objects
is counteracted by the overmerging of others. That is, many close
projections of unrelated objects are counted as one object. This
effect can be quantified by simulations.
Second, because the artificial splitting of large
galaxies into several smaller ones preserves the total luminosity, the 
effect shifts the number counts both vertically and horizontally in
such a way that the predictions of models are not greatly affected. 
Figure \ref{figcounts} shows the effect of overcounting for a non-evolving
model with $q_0 = 0.01$. We have assumed here that galaxies at $\I = 22$
are counted on average as three objects, and that the overcounting
decreases with magnitude such by $\I = 30$ objects are counted only
once.

\subsection*{``Total'' magnitudes}
Different photometry packages use different techniques to estimate
total magnitudes. These estimates are based on extrapolations of galaxy
sizes or surface brightnesses below the initial isophotal limits of
detection, and face serious difficulties in the particular cases of
overlapping galaxies. Even for isolated galaxies, the reliance on the
initial estimates for galaxy sizes
and/or surface brightness profiles can introduce serious biases
for galaxies near the detection threshold.
Nevertheless, before addressing the potential biases,
it is interesting to compare the results from the different
catalogs.

Brighter than $\V = 28$, magnitudes from the different catalogs agree
reasonably well.  Even if the photometry were in principle perfect, the
different splitting algorithms would introduce some scatter at bright
magnitudes.  Figure \ref{figmag} shows comparisons between the
\citeN{WBDDFFGGHKLLMPPAH96}, \citeN{MSCFG96}, and Couch (1996)
magnitudes.  The total magnitudes agree to within 0.5 mag rms over the
full magnitude range. There is a systematic trend for the Metcalfe et
al.  magnitudes to be brighter than the Williams et al. mags. The
difference increases toward faint mags, and is as much as 0.3 mag at
$\V =28$.  The systematic differences between Couch and Williams et al.
catalogs are less than 0.15 mag. The good agreement between catalogs
suggests that the different photometry packages all do more-or-less
the same thing. The discussion of the selection biases below,
which focuses specifically on the \citeN{WBDDFFGGHKLLMPPAH96} catalog,
is thus likely to apply to all the catalogs.

\section*{Galaxy radii}
Figure \ref{figradii} shows the distribution of first-moment radii of galaxies
from the Williams et al. HDF catalog and the Medium Deep Survey
\cite{RGO97}.\footnote{
The Medium Deep Survey (MDS) catalog is based on observations with the
NASA/ESA Hubble Space Telescope, obtained at the Space Telescope Science
Institute, which is operated by the Association of Universities for
Research in Astronomy, Inc., under NASA contract NAS5-26555.  The
Medium-Deep Survey is funded by STScI grant GO2684.}
These isophotal radii keep
decreasing right down to the detection limits of the survey.
At $\V = 27$ the typical first-moment radius corresponds to
less than 2 kpc for {\it for galaxies at any redshift}, for cosmologies
with $q_0 > 0$. The first-moment radius is defined as
\begin{equation}
r_1 = \sum r I(x,y) / \sum I(x,y),
\end{equation}
where $I(x,y)$ is the intensity in each pixel. The quantity 
$R_k = 2 r_1$ is often referred to as the ``Kron radius''.
\citeN{Kron78} showed that in typical ground-based surveys the radius
$R_k$ typically encompasses 90\% of the light from an object.
If the intensity
could be measured precisely over the entire galaxy, the relations
between first-moment radii and scale radii would be 
$r_1 = 2 \alpha = 1.19 r_e $  for spirals, and 
$r_1 = 2.28 r_e $ for ellipticals. The size--magnitude relations
for non-evolving $L^*$ ellipticals and spirals are shown in Fig.
\ref{figradii}, for 
two values of $q_0$.

Taken at face value, the steep radius--magnitude relation implies that
faint galaxies are intrinsically very compact. However, $r_1$ is 
computed using pixels above the isophotal threshold, and thus is
subjected to severe biases as we approach the limiting magnitude
of the survey. These biases enter in several ways:
(1) because of isophotal selection, at fixed total magnitude
larger galaxies will preferentially disappear from the sample; (2)
near the detection limit of the survey the first-moment radius 
$r_1$ becomes progressively biased toward smaller values; and (3) 
near the detection limit, total magnitudes computed either from
Kron magnitudes or from the flux within some multiple of the isophotal
area become progressively biased toward faint values.
While effects (2) and (3) can be partially controlled by measuring 
\citeN{Petrosian76} radii and magnitudes, isophotal selection
effects inevitably bias the sample of galaxies selected for such
measurements.

We can quantify these selection effects by creating
simulated images with the same noise properties as the HDF
images and running them through the same source detection and photometry
routines. The best way to do this would be to add a small number of
galaxies to the HDF image itself and reprocess it hundreds of times.
However, for the purpose of this conference I have taken the shortcut
of constructing only two simulated images, one with face-on exponential
disks and the other with deVaucouleurs profiles. The galaxies have
magnitudes between 25 and 30 and a range of scale lengths. The images
are somewhat less crowded than the real HDF, so crowding should not 
affect the results much. In any case, the results here are meant to 
be indicative rather than definitive. The noise was simulated as described
by \citeN{FB98}, and the source detection and photometry
were done using the same version of FOCAS with the same parameters 
used by \citeN{WBDDFFGGHKLLMPPAH96}.

Figure \ref{figcompre} 
shows the selection boundary for galaxies with exponential profiles
(left panel) and deVaucouleurs profiles (right panel) . The numbers in
the figure give the fraction of the input sample of galaxies recovered
as a function half-light radius $r_e$ and total magnitude, $\I$. The
solid curves show the limits of the survey expected from the rms sky
noise of the HDF images, assuming FOCAS detects sources with AB
magnitude $\I < 30.6$ within a radius of $0.064"$. This seems to be a
reasonably good model of the 50\% completeness limit of the Williams et
al. catalog.  To put this in context, the dashed curves show the
size--magnitude relation for non-evolving $L^*$ galaxies. A typical
present day spiral would drop below the selection limits of the survey
by $z = 1.2$, even though its total magnitude is brighter than $\I =
28$ out to $z=2.5$.  A typical elliptical would be invisible beyond
$z=1.8$.

Figure \ref{figcompre} 
illustrates the point that the HDF and other deep surveys
should not be treated as flux-limited surveys. The sizes of galaxies
matter as much as their total luminosities. This fact, although
well known, is often ignored in comparing model counts to the data.
With the HDF and future surveys it is extremely important to take this
into account.

\subsection*{Robust constraints on galaxy sizes}
Of course, it would be nice if we could infer something robust
about the physical sizes of galaxies in the HDF, independent of
any assumed model. To do this, we must translate the selection boundaries
in Figure 5 from the theoretical $M_{\rm tot}, r_e$ plane, to the 
observed $M_{\rm iso}, r_1$ plane. This can be done analytically or
numerically for the different profiles, and can also be done 
empirically from the FOCAS measurements of the simulated galaxies. 
Figure \ref{figcompr1} 
shows the translated curves from Fig. \ref{figcompre}, along with the
observed distribution of sizes and magnitudes of galaxies in the
HDF. The thin lines show the fiducial galaxies from Fig. \ref{figcompre}. 

Brighter than $\I = 27.5$, the locus of points clearly peaks at values
of $r_1$ that are well away from the selection boundary. Hence it
is probably safe to assume that the relatively compact sizes and
high surface brightnesses of the HDF galaxies are a real phenomenon,
and not an artifact of galaxy selection. The apparent sizes are smaller
than those of present-day $L^*$ disk-dominated spirals, 
but not significantly smaller
than those of luminous ellipticals.  Very little can be said about
the intrinsic sizes of galaxies fainter than $\I = 27.5$, because it
it likely that many galaxies are missed
due to low surface brightness. The  survey limit determines the 
upper bound on $r_1$, and the PSF determines the lower bound. 
This truncation of the survey by the isophotal limit means that
galaxy counts fainter than $\I = 27.5$ must also be viewed 
as a lower limit. Also large corrections are required
to translate from FOCAS isophotal to true total magnitudes near
the survey limit. Such corrections can be derived for models, but 
cannot be derived with any certainty from the data themselves.

\section*{Lyman Break Galaxies}

The past two years have brought an explosion in the number of
star-forming galaxies identified at redshifts $z > 2$.  The HDF has
contributed to this by providing a set of robust high-redshift galaxy
candidates several magnitudes below the detection limits of current
Keck spectroscopy.  \citeN{MFDGSF96} used the statistics of
Lyman-break objects in the HDF to estimate the luminosity-density in
redshift intervals centered at $<z> = 2.75$ and $<z>=4$. Together with
constraints from the CFRS survey \cite{CFRS1} and a local H$\alpha$
survey \cite{GZAR95}, this analysis provides an indication of the
metal-formation rate and integrated star-formation rate as a function
of redshift. Because it puts the observations in a physical context,
the ``Madau diagram'' has become a popular foil for discussing and
testing models of galaxy formation, and
has been subject to 
a fair amount of scrutiny. Much of the debate centers on how
much dust there is in star-forming galaxies at high redshift, 
and on what corrections are needed for extinction and for galaxies
that are too dusty to detect (see contributions by 
Madau, Rowan-Robinson, and Meurer in this conference).
The luminosities of the $<z> = 2.75$ and the $<z> = 4$ samples
are computed at essentially the same rest-frame wavelengths 
$(\sim 1500$ {\AA}), and thus are probably subject to the similar
amounts of extinction. If anything the extinction should be less
on average for the $<z> = 4$ sample, since there is likely to be
less dust at high redshift. This suggests that the decreasing luminosity 
density from $z = 2.75$ to $z = 4$ is a real effect, indicative
of a change in the overall star-formation rate of the universe.
 
However, it is important to keep in mind that the Lyman break objects,
like all other galaxies in the HDF, are selected above a fixed
isophote, and are subject to the same selection effects. For example,
at fixed total apparent magnitude and physical scale length, a galaxy
at $z=4$ will have a mean surface brightness 1.25 mag fainter than a
galaxy at $z=2.75$.  If the Lyman-break objects fall near the selection
boundary, this change in surface brightness could in principle
translate to a 0.5 dex change in the inferred luminosity density, which
is roughly what is observed between these two redshifts.

To explore this effect, we have examined the statistics of Lyman
Break galaxies in the models of \citeN{FB98}, comparing
what would be derived from an ideal survey, to what we derive from
simulated images of the HDF. Figure \ref{figdropt} shows position
of model galaxies in \citeN{MFDGSF96} color--color plot. For this
figure, the total magnitudes of the galaxies from the models are used,
and we have truncated the sample at $\I = 27.6$, which corresponds
to the $\sim 10 \sigma$ detection limit of the HDF. The 
\citeN{MFDGSF96} selection criteria applied to this sample yields 
a total of 62 ``U dropouts'' and 232 ``B dropouts'' (compared
to 69 and 14, respectively, in the real HDF).

Figure \ref{figdrop} shows FOCAS measurements {\it of the same 
model} from simulated images. This figure should be directly 
comparable to those in \citeN{MFDGSF96}. The color selection
criteria yield 10 U dropouts and 21 B dropouts. 
This illustrates the potential severity of the selection effects on the
statistics of Lyman break objects. The number density of Lyman-break
objects in the input model differs from the number recovered by FOCAS
by roughly one order of magnitude! 

The effect on the inferred luminosity density is much less severe, but
is still quite significant, at least for the models we have tested.
Table 2 shows the total magnitude (from the summed flux of all the
detected galaxies) and derived metal-formation rates $\dot{\rho}_Z$ (in
units of $\rm M_\odot\,yr^{-1}\,Mpc^{-3}$) for Lyman-break galaxies
selected directly from the model, from the simulated images, and from
the real HDF. Isophotal selection decreases the inferred
metal-formation rate by a factor of 1.5 for the U dropouts and 4.7 for
the B dropouts. However, the high metal-formation rate at $z>3.5$  for this
model is still at odds with the data (see \citeNP{FB98}). Tests with another model show similar results, although the
correction factors factors to go from the FOCAS-measured luminosity
density to the luminosity density in the underlying model are not
the same.

\section{Summary}
I have tried in this presentation to give some quantitative insight 
into how selection effects influence the interpretation of the HDF data.
In general, the treatment of the HDF as a flux limited survey is
a bad approximation. Magnitudes and radii of galaxies are strongly
affected by the measurement techniques (which vary from catalog to
catalog) and by the detection algorithm (which is virtually the same
in all catalogs generated to date). Faintward of $\I = 27.5$, the angular
size distribution of galaxies in the HDF is severely truncated by
the survey selection boundaries. Brighter than this, it appears that
the HDF galaxies are in general more compact than $L^*$ spirals, but
not necessarily more compact that luminous present-day ellipticals.
The angular-size distribution is reasonably well matched by a 
low-$q_0$ pure-luminosity evolution model \cite{FB98},
although such a model does not reproduce the statistics of Lyman-break 
objects, even when selection effects are accounted for.

The survey selection criteria can have a very strong effect on the 
statistics of Lyman-break objects, potentially influencing the 
source counts by an order of magnitude, and the inferred luminosity
density by up to a factor of 5. The corrections factors
depend on the input model. 

While this is a fairly discouraging view of quality of information
on faint galaxies that can be gleaned from the HDF, nevertheless
I think it is fairly accurate. Approximate corrections for some of
these selection effects can be derived from the HDF data themselves,
but for the most part the magnitude of the selection bias depends
on the intrinsic distributions of galaxy sizes and luminosities, which
are not well constrained at faint magnitudes. 
Fortunately, the selection effects are fairly straightforward to model. 
Comparisons of galaxy evolution models to the HDF really must involve this step.

\acknowledgments

This work was supported by NASA grant AR-06337, awarded by the Space
Telescope Science Institute, which is operated by the Association
of Universities for Research in Astronomy, Inc., for NASA under
contract NAS5-26555.

\bibliography{apjmnemonic,bib} 
\bibliographystyle{apj}

\def\~{\thinspace}

\begin{deluxetable}{ll}
\small
\tablewidth{0pc}
\tablecaption{Selected HDF Catalogs}
\tablecolumns{6}
\tablehead{
\multicolumn{1}{l}{Catalog}&
\multicolumn{1}{c}{Comments}}
\startdata
Couch & Sextractor  \nl
	     & 1683 Objects on WF chips, selected in F814W. \nl
	     & Isophotal limit $\mu_{814} = 26.1$ in 0.016 arcsec$^2$ \nl
Lanzetta et al. & Sextractor \nl
		& 1925 Objects on WF chips, selected in separate bands \nl
		& Isophotal limit $\mu_{814} = 26.2$ in 0.048 arcsec$^2$ \nl
Williams et al. & modified FOCAS \nl
		& 3086 Objects selected in F814W+F606W summed image \nl
		& Isophotal limit $\mu_{814} = 27.5$ in 0.04 arcsec$^2$ \nl
Metcalfe et al. & Detection algorithm described in Metcalfe et al. 1991\nl
		& ~3700 Objects \nl
		& Thresholds not specified, but probably the deepest. \nl

\enddata
\end{deluxetable}

\def\~{\thinspace}

\begin{deluxetable}{lllllll}
\small
\tablewidth{0pc}
\tablecaption{Model vs. Data for HDF Lyman Break Objects}
\tablecolumns{6}
\tablehead{
\multicolumn{1}{l}{}&
\multicolumn{1}{c}{Number of}&  
\multicolumn{1}{c}{Total}&  
\multicolumn{1}{c}{}&  
\multicolumn{1}{c}{Number of}&  
\multicolumn{1}{c}{Total}&  
\multicolumn{1}{c}{} \nl 
\multicolumn{1}{l}{Data set}&
\multicolumn{1}{c}{U dropouts}&  
\multicolumn{1}{c}{$\V$}&  
\multicolumn{1}{c}{$\dot{\rho}_Z$}&
\multicolumn{1}{c}{B dropouts}&  
\multicolumn{1}{c}{$\I$}&  
\multicolumn{1}{c}{$\dot{\rho}_Z$} \nl 
}
\startdata
Input model & 62 & 21.15 & $4.9 \times 10^{-4}$ & 232 & 19.14 & $50 \times 10^{-4}$ \nl
FOCAS & 10 & 21.58 &       $3.3 \times 10^{-4}$ & 21 & 20.83 & $10 \times 10^{-4}$ \nl
HDF & 69 & 21.48 &         $3.6  \times 10^{-4}$& 14 & 23.29 & $1.1 \times 10^{-4}$ \nl

\enddata
\end{deluxetable}

\begin{figure}[p]
\epsscale{0.7}
\caption{A portion of the HDF, with objects brighter than $\V = 28.5$ 
from the Couch (1996) catalog marked. Note that there are
quite a lot of sources clearly visible below this magnitude limit, many
of which appear in the catalogs, but we have not marked them on
the images to avoid clutter. ``Total'' F606W AB magnitudes are marked 
for a few galaxies to give an indication of the agreement between catalogs.
}
\end{figure}

\begin{figure}[p]
\epsscale{0.7}
\caption{Same for the Lanzetta et al. (1996) catalog.}
\end{figure}

\begin{figure}[p]
\epsscale{0.7}
\caption{Same for the Williams et al. (1996) catalog.}
\end{figure}

\begin{figure}[p]
\epsscale{0.7}
\caption{Same for the Metcalfe et al. (1996) catalog.}
\end{figure}

\begin{figure}[p]
\centerline{\plotone{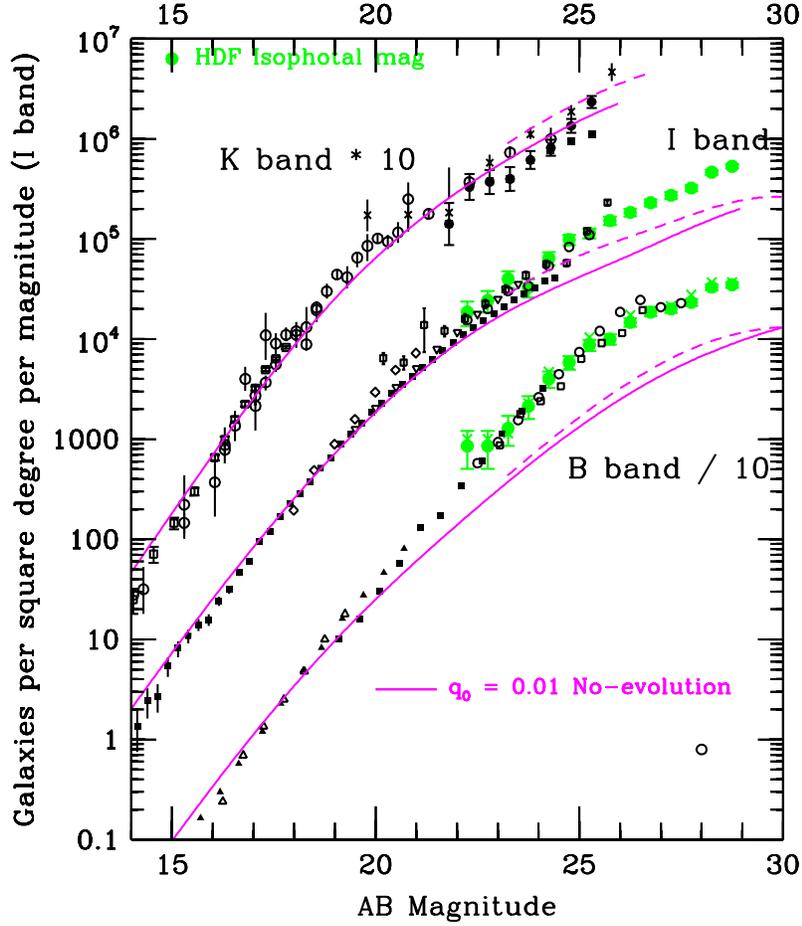}}
\caption{\label{figcounts}
HDF counts from Williams et al. (1996), together with a compilation
of counts from ground-based surveys. The solid curves show a 
non-evolving model with $q_0 = 0.01$. The dashed curves show the
same model, but with overcounting included. For magnitudes $22 < m_{AB} < 30$,
the counts have been multiplied by $1+(30-m)/4$, and the 
magnitudes have been increased by $2.5 \log(1+(30-m)/4)$. 
This illustrates the fairly modest affect that splitting and merging
algorithms have on the ability to model galaxy number counts.
}
\end{figure}

\begin{figure}[p]
\centerline{\plottwo{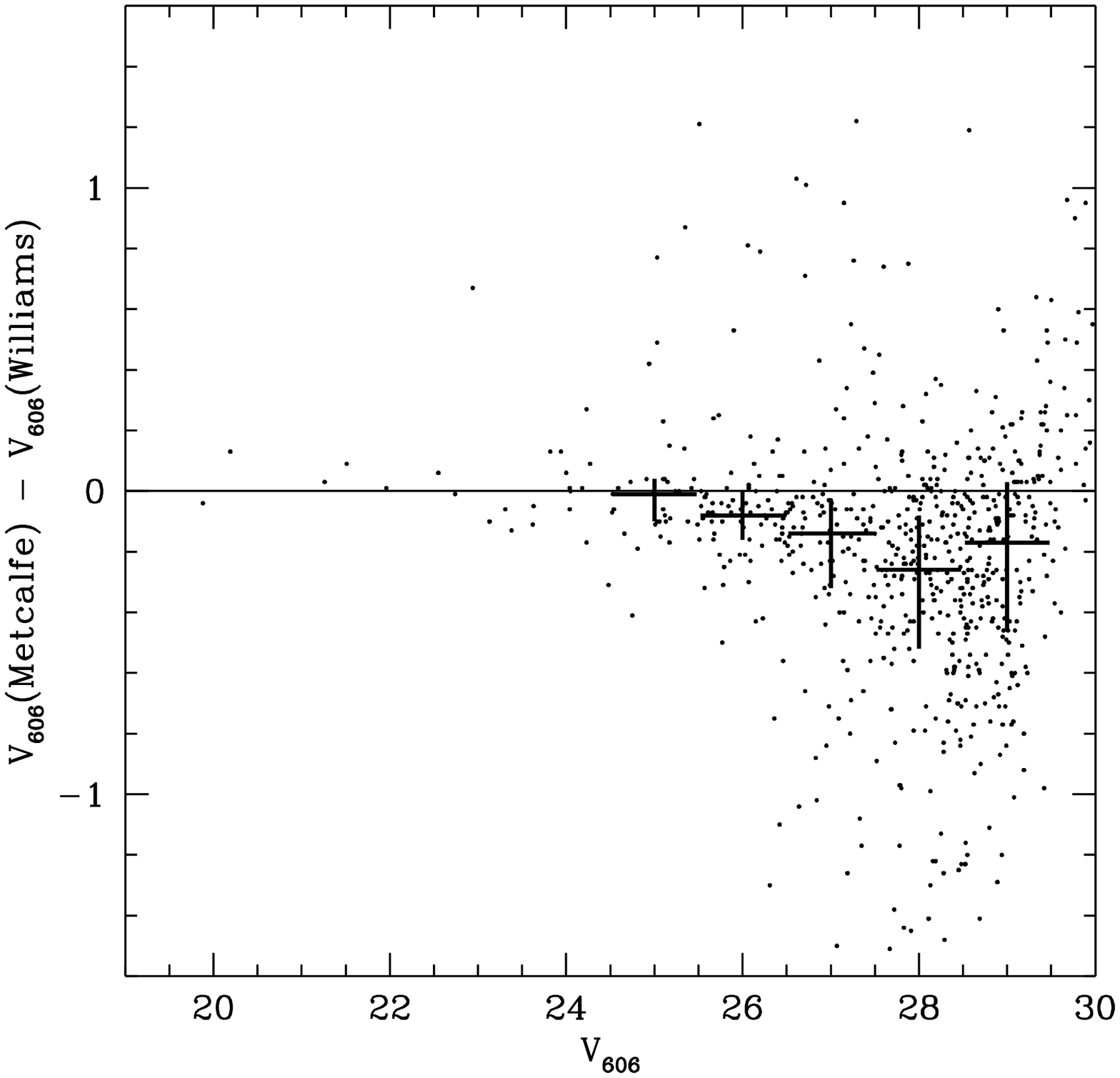}{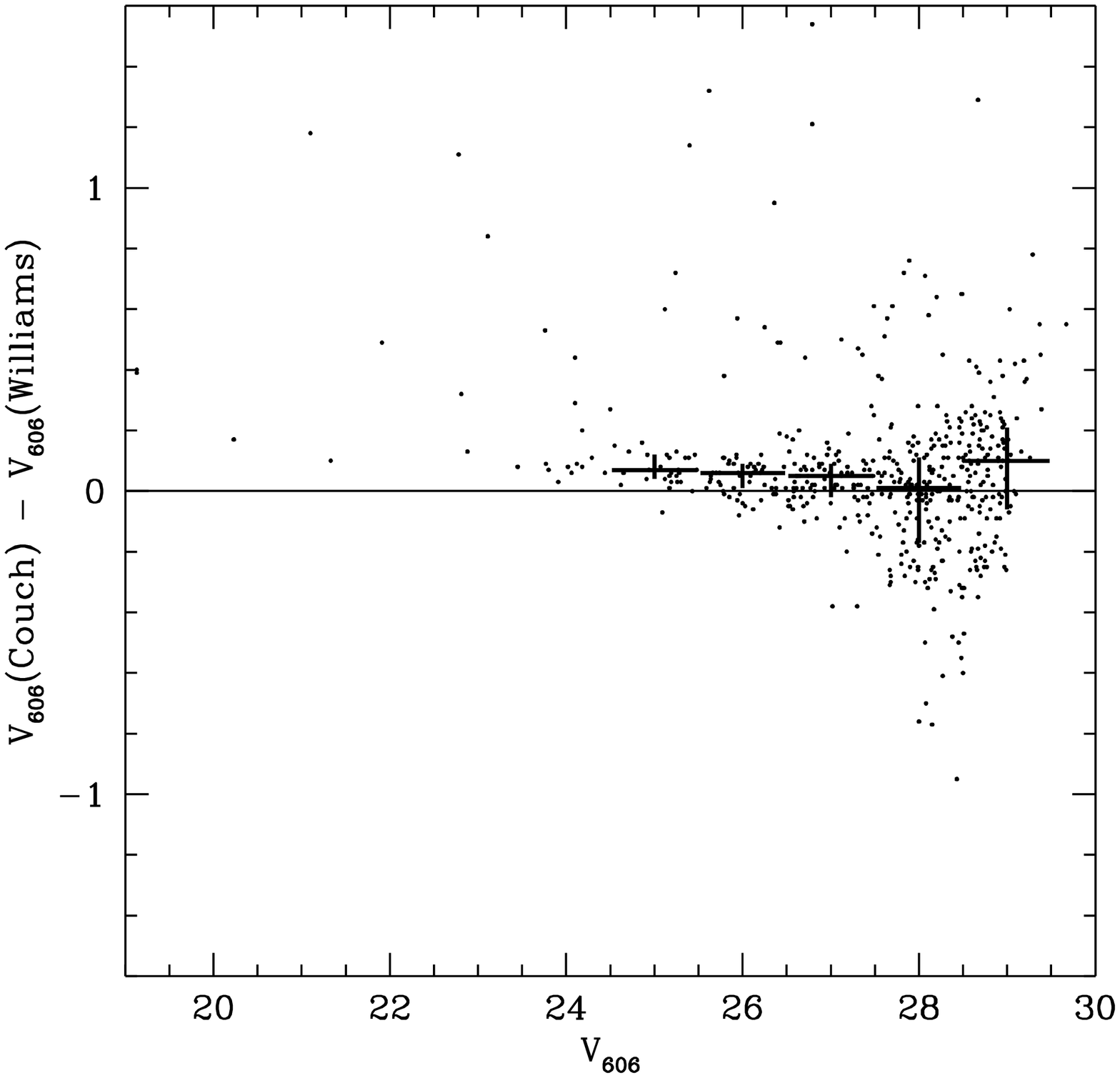}}
\caption{\label{figmag}
The left panel shows a 
comparison of $\V$ total magnitudes from the Metcalfe et
al. and Williams et al. HDF catalogs for galaxies on WF2. The 
points with error bars show the median and first and third quartiles
of the distribution of the residuals in intervals of 1 magnitude.
The right panel compares the Williams et al. and the Couch 
magnitudes.
}
\end{figure}

\begin{figure}[p]
\centerline{\plotone{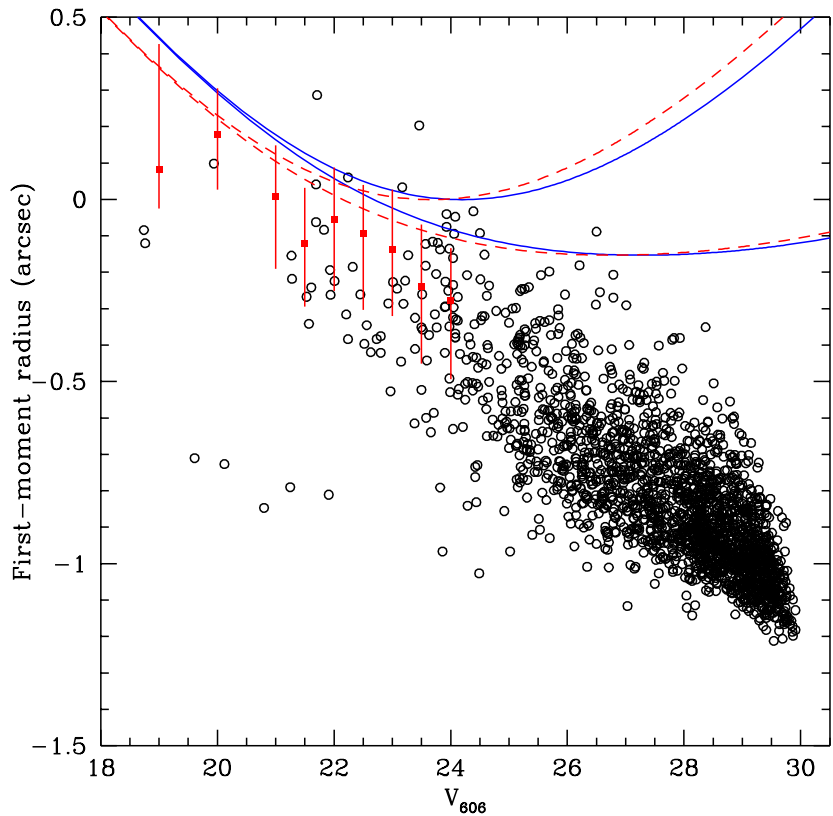}}
\caption{\label{figradii}
Radius--magnitude relations for galaxies in the HDF
Williams et al. (1996) catalog (points) and the MDS (points
with error bars). The MDS distribution was derived from the catalogs
on the MDS web site. The curves show the behavior for non-evolving
$L^*$ spirals (solid curves) and ellipticals (dashed curves). The
upper pair of curves is for $q_0 = 0.5$; the lower is for $q_0 = 0.01$.
{\it These curves do not take into account selection biases.} The severity
of the biases can be seen in the next two figures.
}
\end{figure}

\begin{figure}[p]
\centerline{\plottwo{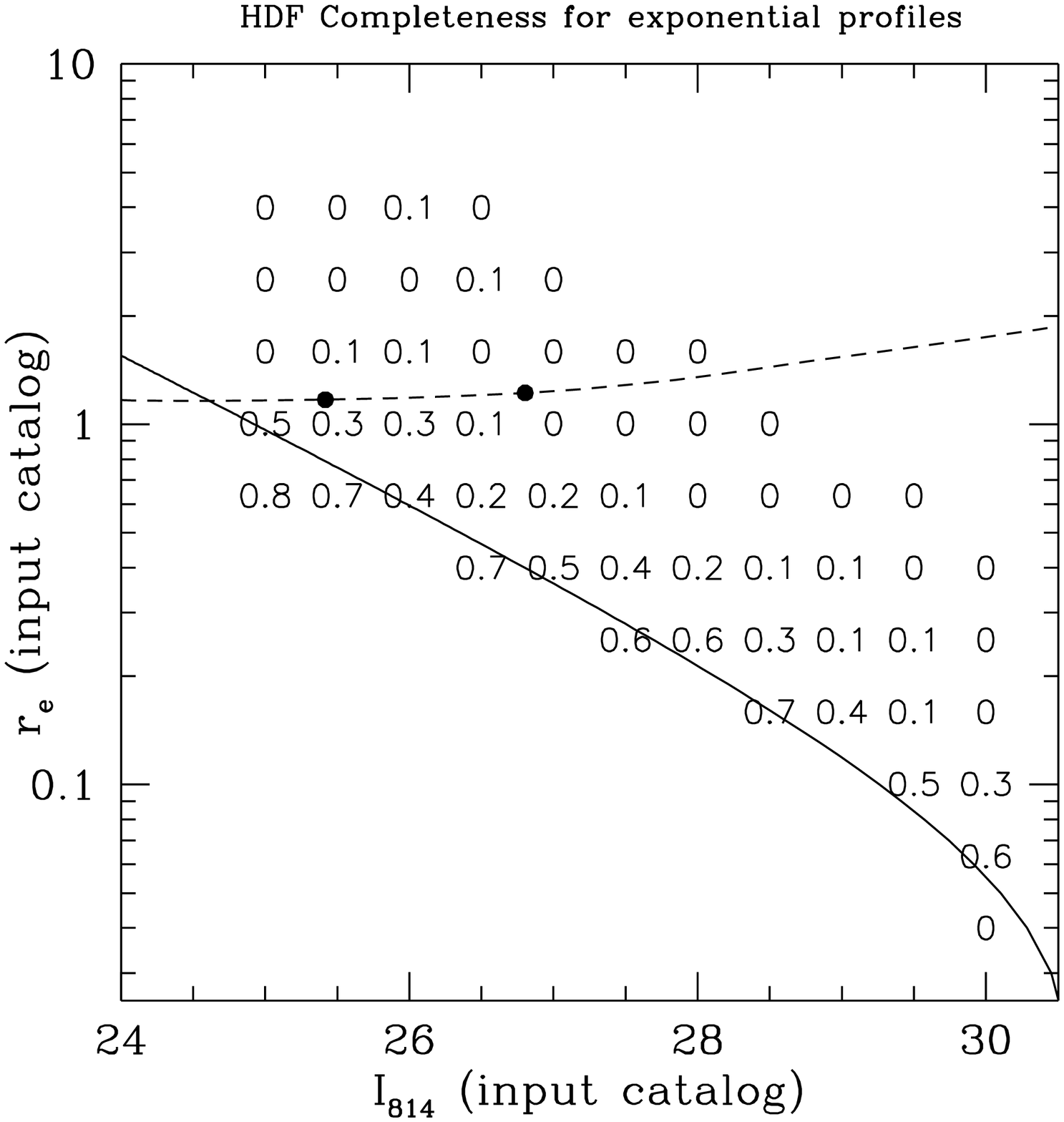}{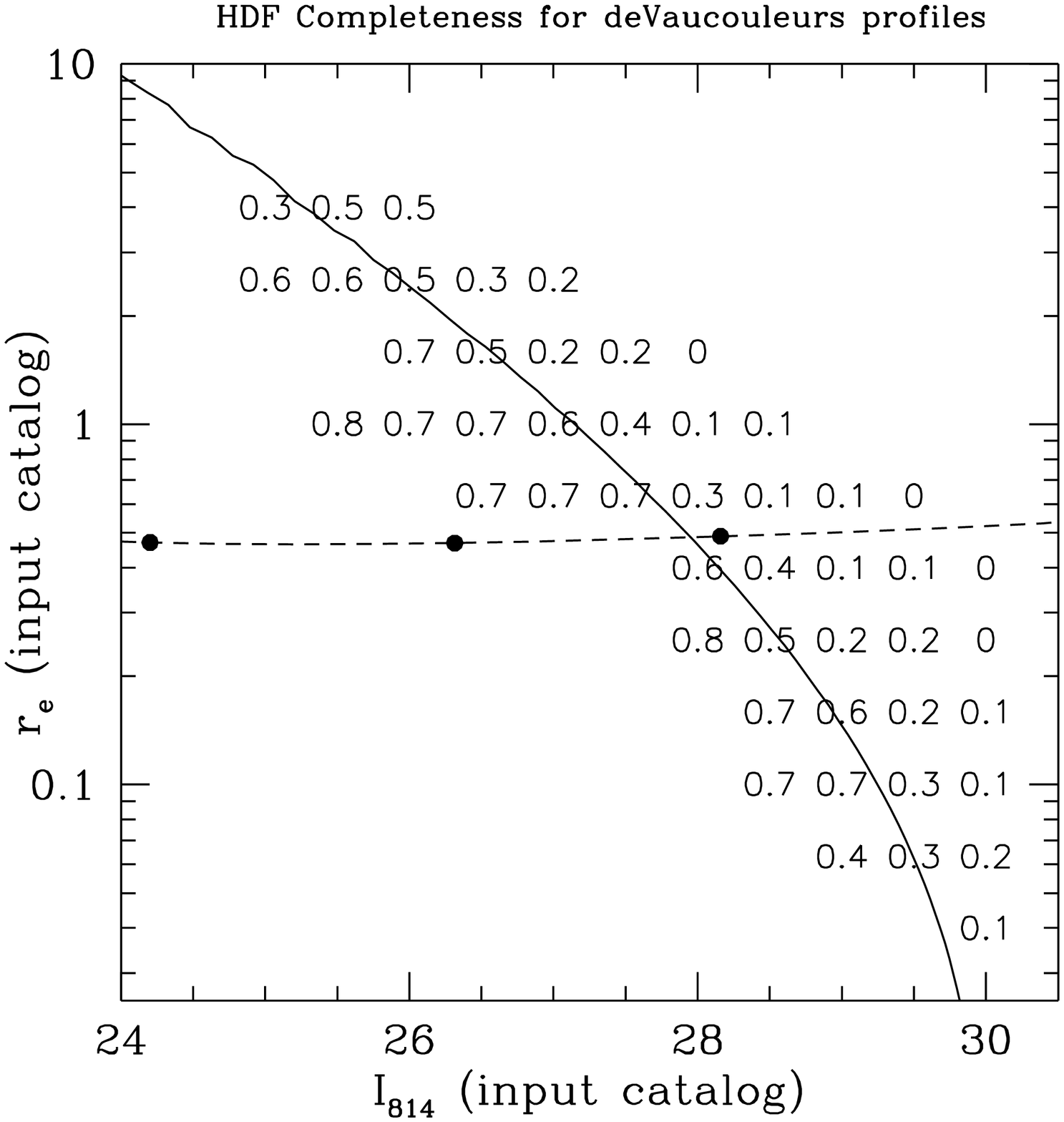}}
\caption{\label{figcompre}
Empirical selection boundary for the HDF as a function of 
half-light radius and total $\I$ magnitude input to the simulations.
The left panel shows the limits for face-on galaxies with pure
exponential profiles. The right panel shows the same for face-on
galaxies with $r^{1/4}$ deVaucouleurs profiles.
As there were typically 20 galaxies per bin in the input sample, 
these estimates are subject to small-number statistics. 
The solid curves show an analytical model of the FOCAS selection 
criteria, where we have assumed that a galaxy will be detected
if it is brighter than $\I=30.6$ within a radius of 0.064 arcsec. 
The dashed curves show the relation between $r_e$ and $\I$ for
non-evolving galaxies with $M_B = -21.1$, for $H_0 = 50
\rm \,km\,s^{-1}\,Mpc^{-1}$, and $q_0 = 0.5$. We have assumed a
scale-length $\alpha = 6 \rm \,kpc$ for the spiral, and $r_e = 4 \rm \,kpc$
for the elliptical, and have adopted spectral-energy distributions
from Coleman, Wu, \& Weedman 1980. 
The solid dots on the curves for
both galaxies correspond to $z = 1, 1.5$, and 2 (from left to right);
the $z=1$ point for the spiral is just off the left of the plot at 
$\I=23.7$.
}
\end{figure}
\nocite{CWW80}

\begin{figure}[p]
\centerline{\plotone{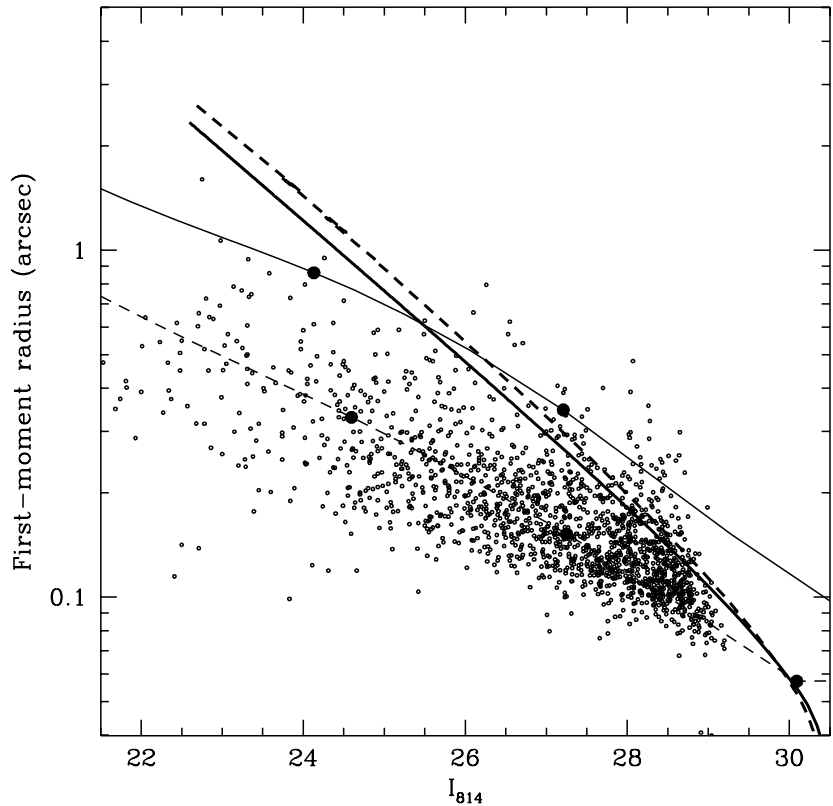}}
\caption{\label{figcompr1}
HDF angular-size -- magnitude relation. Sizes are the first-moment
radii and magnitudes are FOCAS $\I$ isophotal magnitudes.
The thick lines show the selection boundaries for spirals (solid line)
and ellipticals (dashed line), with the same assumptions used
for Figure 4. The thin lines show the fiducial non-evolving
$L^*$ spiral and elliptical galaxies. 
The solid dots on the curves for
both galaxies correspond to $z = 1, 1.5$, and 2 (from left to right; 
the first dot for the spiral is off the plot).
}
\end{figure}

\begin{figure}[p]
\centerline{\plottwo{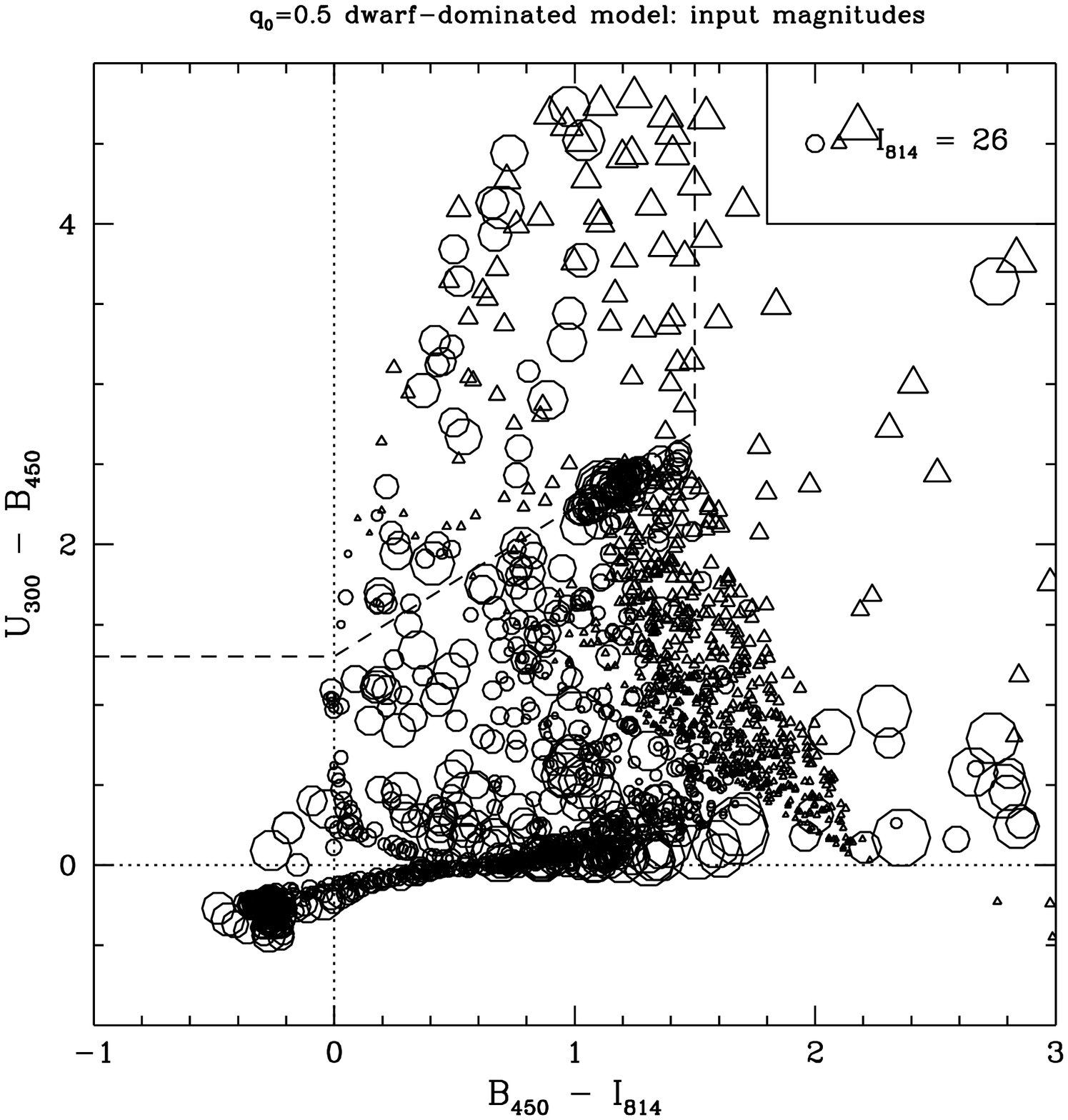}{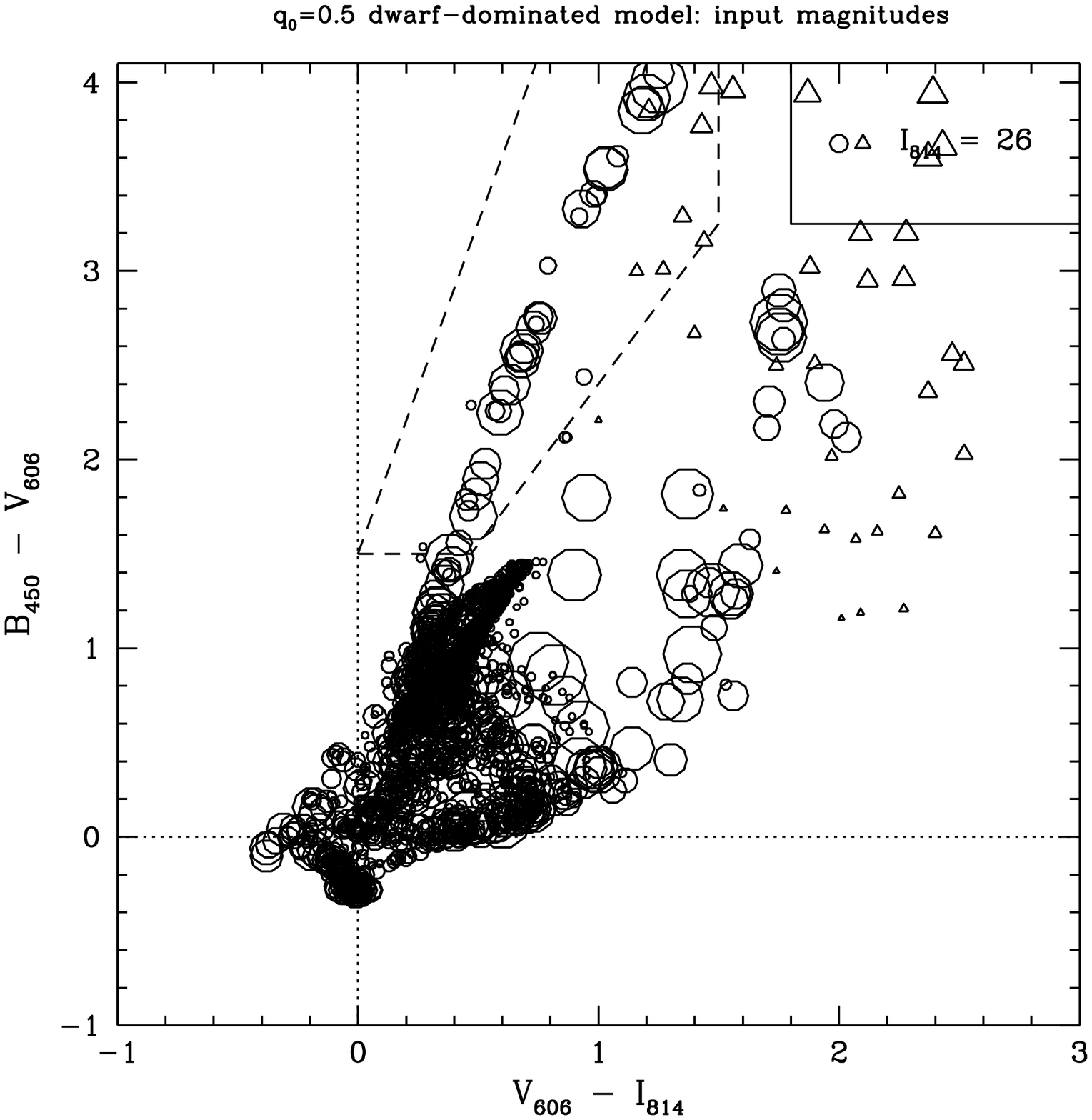}}
\caption{\label{figdropt}
Color-color diagrams fashioned after Madau et al. for galaxies
in the $q_0 = 0.5$ dwarf-dominated model of Ferguson \& Babul 1998.
The positions of the galaxies in this diagram are determined from 
the {\it true total magnitudes} of the input model. The number of
Lyman-break objects within the dashed boundaries should be compared
to the next figure, where the same galaxies have been placed in
simulated images and measured using FOCAS with the Williams et al.
(1996) parameters.}
\end{figure}

\begin{figure}[p]
\centerline{\plottwo{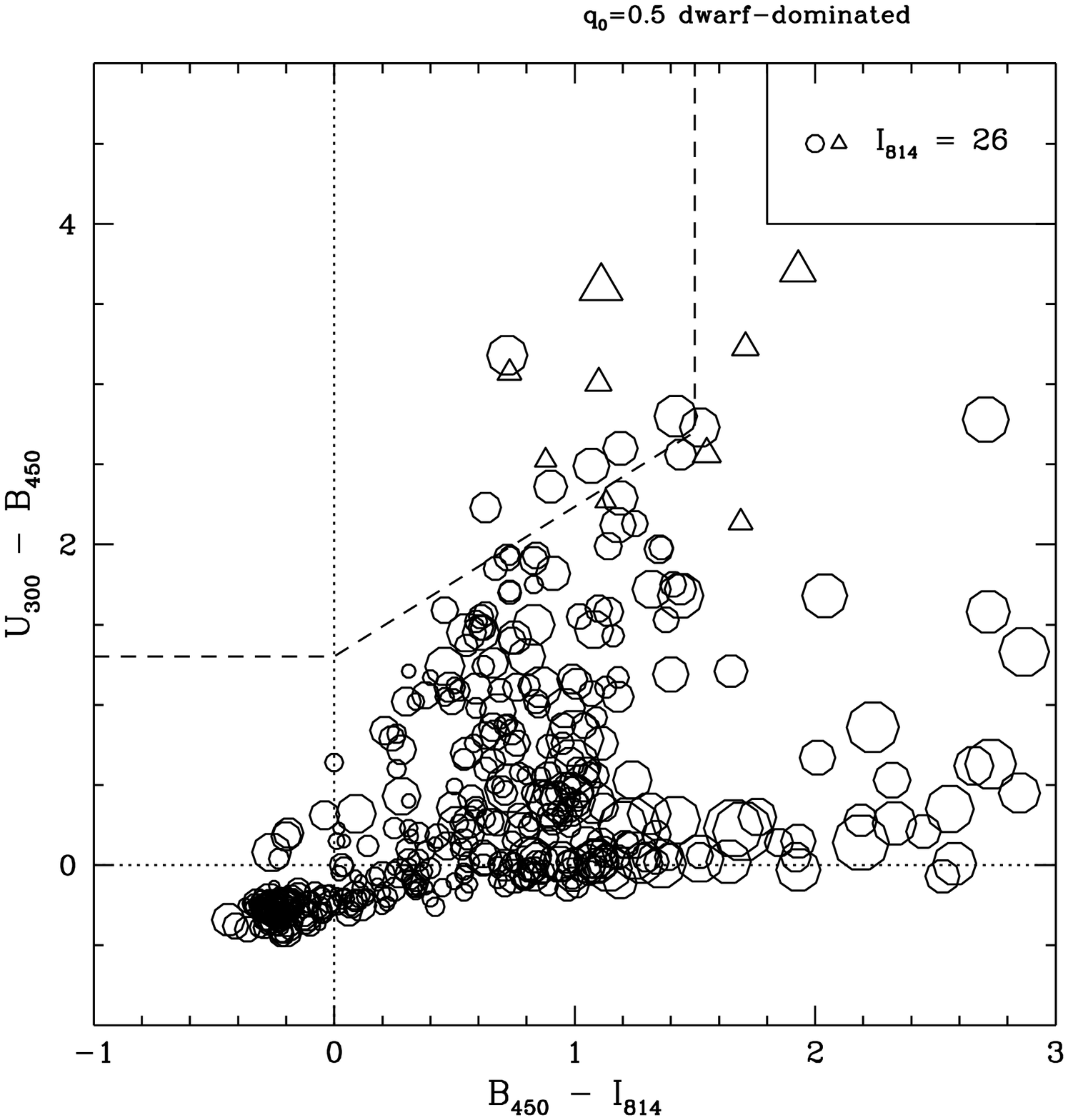}{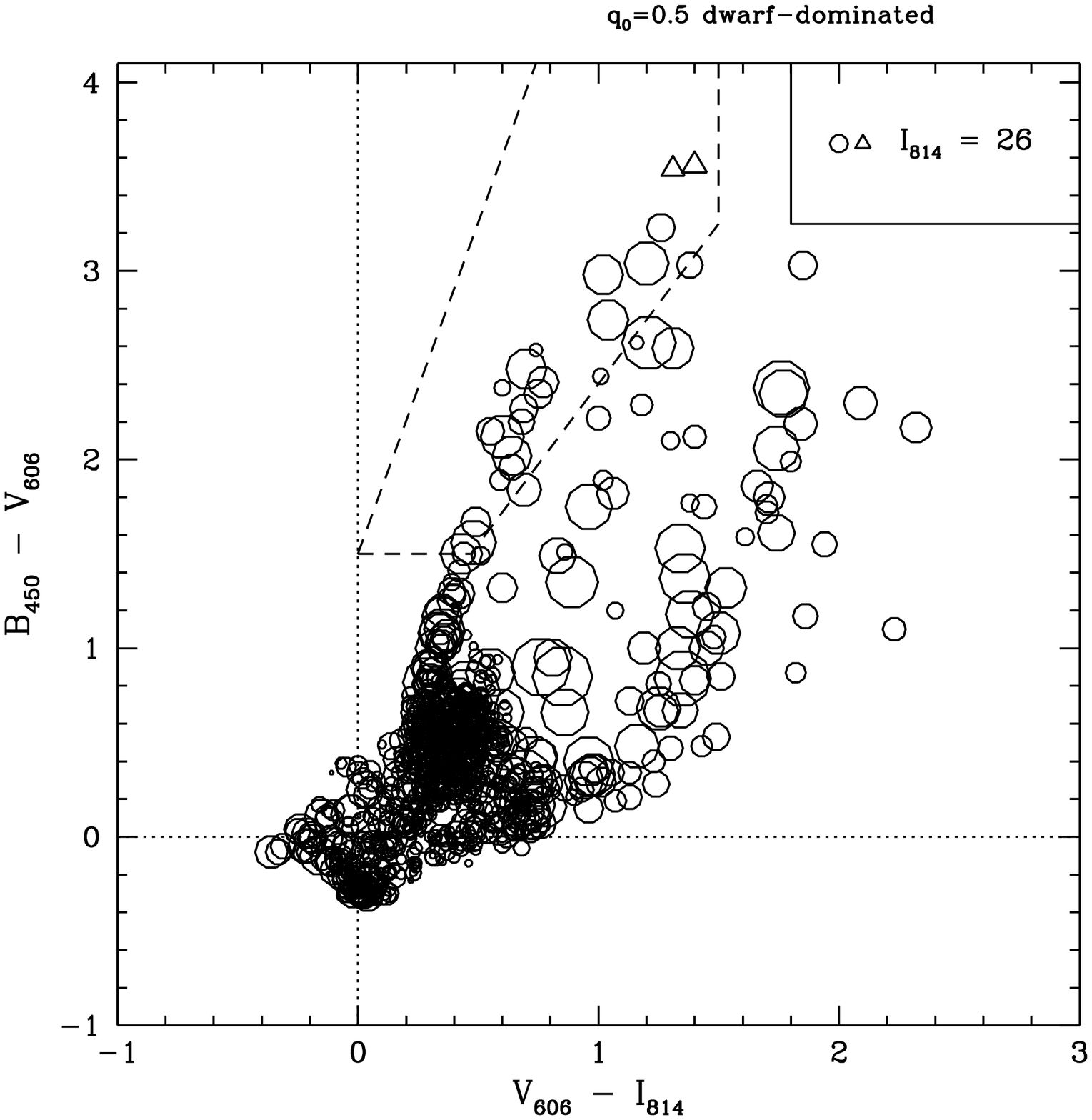}}
\caption{\label{figdrop}
Same as the previous figure, but for galaxies measured using
FOCAS (with the same parameters used by Williams et al.) in simulated
HDF images. Only about 10\% of the Lyman-break objects in
the input model are recovered by FOCAS. There are various reasons
for this. Many of the galaxies that look bright in Fig. 10 have 
large scale lengths, and hence have FOCAS isophotal magnitudes that
are much fainter than their model total magnitudes. In other cases,
the FOCAS photometric errors change the colors enough to move the
galaxy outside the selection boundary.
}
\end{figure}

\end{document}